\documentclass{IEEEcsmag}

\usepackage{hyperref}
\usepackage{upmath}
\usepackage{xspace}
\usepackage{xcolor, soul}
\usepackage{ragged2e}
\usepackage{anyfontsize}
\usepackage{subcaption}
\usepackage{booktabs}

\jvol{XX}
\jnum{XX}
\paper{8}
\jmonth{July}
\jname{Computer}
\pubyear{2021}

\newtheorem{example}{Example}

\newcommand\CAISs{CAISs\xspace}
\newcommand\CAIS{CAIS\xspace}

\begin{document}
\bstctlcite{IEEEexample:BSTcontrol}
\sptitle{Department: Head}
\editor{Editor: Name, xxxx@email}

\title{Collaborative AI Needs Stronger Assurances Driven by Risks}

\author{Jubril Gbolahan Adigun}
\affil{University of Innsbruck, Austria}

\author{Matteo Camilli}
\affil{Free University of Bozen-Bolzano, Italy}

\author{Michael Felderer}
\affil{University of Innsbruck, Austria and Blekinge Institute of Technology, Karlskrona, Sweden}

\author{Andrea Giusti}
\affil{Fraunhofer Italia Research, Bolzano, Italy}

\author{Dominik T Matt}
\affil{Free University of Bozen-Bolzano and Fraunhofer Italia Research, Bolzano, Italy}

\author{Anna Perini}
\affil{Fonzazione Bruno Kessler, Trento, Italy}

\author{Barbara Russo}
\affil{Free University of Bozen-Bolzano, Italy}

\author{Angelo Susi}
\affil{Fondazione Bruno Kessler, Trento, Italy}

\markboth{Department Head}{Paper title}

\begin{abstract}

\justify  
Collaborative AI systems (\CAISs) aim at working together with humans in a shared space to achieve a common goal. 
This critical setting yields hazardous circumstances that could harm human beings. 
Thus, building such systems with strong assurances of 
compliance with requirements, 
domain-specific standards and regulations is of greatest importance.
Only few scale impact has been reported so far for such systems since much work remains to manage possible risks.
We identify emerging problems in this context and then we report our vision, as well as the progress of our multidisciplinary research team composed of software/systems, and mechatronics engineers to develop a risk-driven assurance process for \CAISs.
\end{abstract}

\maketitle

\newpage
\section{Introduction} 

As Artificial intelligence (AI) technologies continue to grow and evolve, Cyber-physical Systems (CPSs) increasingly include Machine Learning (ML) components even in critical applications, such as collaborative robots.
The latter systems are referred to as Collaborative AI Systems (\CAISs)~\cite{CamilliREFSQ2021} since they work together with humans in a shared physical space to achieve a common goal and they are enabled by ML components to mimic human perception skills (e.g., visual perception, speech recognition) or learn by demonstration to make the robots flexible and able to accommodate changing requirements.

Driven by technical and economic forces, these systems are gaining momentum in industrial manufacturing as they can enable the transition to the new vision of \emph{Industry 5.0}~\cite{EUIndustry5} by putting AI technologies at the service of a human-centred and resilient industry.
CAISs are expected to build and maintain trust between the human and the machine by ensuring a safe interaction in which the machine does not prevail over human needs. 
To this end, they must meet key quality criteria, including appropriate behavior with respect to social rules, 
industry-specific standards and laws. 
Compliance with such requirements heavily depends on the quality of the ML whose typical goal is to learn a behavior from labelled data and generalize it on unlabeled new cases.

\begin{figure*}[bt]
    \centering
    \includegraphics[width=\textwidth]{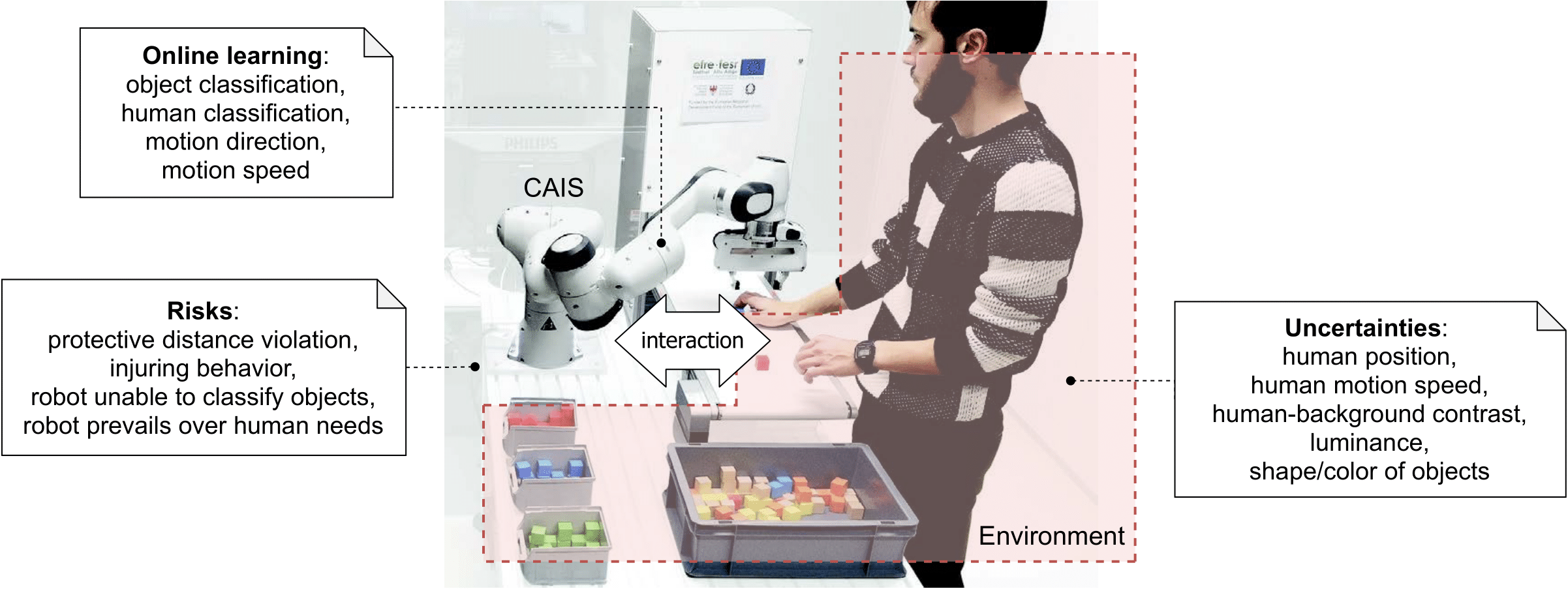}
    \caption{
    ML-equipped CAIS demo running at the ARENA lab., Fraunhofer Italia.}
    \label{fig:running_example}
\end{figure*}

The ML components in \CAISs are embedded in complex and dynamic ecosystems affected by sources of uncertainty that make it difficult to provide strong assurances of compliant behavior.
Figure~\ref{fig:running_example} shows an example of \CAIS in which a robotic arm collaborates with a human operator.
This system has been developed at the ARENA laboratory~\cite{arenaLab} of the Fraunhofer Italia Research institute and it represents a demo for the manufacturing domain. Here, the ML component aims at classifying objects based on shapes and colors and learns from operator's gestures.
The robotic arm and the human being share a physical space and closely collaborate to achieve the common goal of classifying and partitioning the objects passing on the conveyor belt.
During the collaboration, the \CAIS perceives its surroundings and it learns during operation (online), how to classify objects from the human being by demonstration.
The online training may lead to a poor classification accuracy of the ML component.
This in turn can impact the safety of the collaborative application since the robotic arm may move unexpectedly and yield hazards originating from undue physical contacts causing injuries to the human operator.
This specific risk, as well as other risk examples listed in Fig.~\ref{fig:running_example}, are intrinsic to the adoption of ML components whose behavior and performance are referred to as stochastic rather than deterministic.
These aspects are also affected by the many factors of the production environment that are usually partially known during the design stages and may also change (uncertiainties in Fig.~\ref{fig:running_example}).

Several properties, like safety in our example, but also robustness, reliability, and accuracy are crucial for \CAISs.
As highlighted by the recent standards on AI~\cite{isoTR24029,iso79799},
systems enabled by ML components pose new challenges that require systematic engineering processes able to provide strong assurances (i.e., through adequate testing and validation).
As illustrated above, \CAISs yield risks tied to the behavior of the ML components. Understanding these risks is essential for the adoption of these systems in the contexts of collaborative applications regulated by 
regulated by the standard ISO 10218
~\cite{iso51330,iso41571} and enhanced by the ISO/TS 15066~\cite{iso62996}
. To support the existing demand, a step change in engineering approaches and assurance methods is required 
and much work needs to be done to manage possible risks and increase the scale impact of \CAISs
~\cite{HolzingerMLKE2021}.

In this paper, we first discuss the current research landscape by looking backward to relevant topics that have been addressed so far and by looking forward to open issues.
Then, we provide an overview of our risk-driven assurance process, which focuses on: (semi) automated engineering processes for compliance requirements extraction, representation, and traceability;
alternative hypotheses of risk representation and management for \CAISs, and risk-driven assurance methods. We report our vision, the progress, and the next steps of our multidisciplinary research team composed of software/systems, and mechatronics engineers.

\section{Where Are We?}

In this section, we discuss the current research landscape in engineering AI systems in critical domains like \CAISs.
We first look backward--at relevant topics that have been addressed so far--and then forward--at open issues currently limiting scale impact of these systems in a real-world setting.

\subsection{Looking backward}

Recent advances in ML yield opportunities in the robotics domain due to the ability to implement functions we usually associate with human skills, such as perception, inference, planning, and learning activities.
These advances recently brought flexibility in human-robot collaboration by making the robotic systems able to swiftly adapt to different tasks with \emph{imitation learning}~\cite{Giusti2018}.
Furthermore, perception and control functions have been enhanced by Deep Neural Network (DNN) models to implement, for instance, visual perception components able to monitor the surroundings and thus plan accordingly.

The adoption of ML in critical domains elevates concerns on quality aspects, like reliability, safety, accuracy, and robustness, as described in the standards ISO/IEC TR 24029-1~\cite{isoTR24029} on assessment of the robustness of neural  networks, and ISO/IEC DTS 4213~\cite{iso79799} (currently under development) on assessment of ML classification performance.
Assurance is an essential part of putting an ML component into a target system in production.
In critical domains like \CAISs, this plays an important role also for certification purposes.
For instance, collaborative robots~\cite{iso62996} require adequate actions able to assure that the system can justifiably be trusted by the humans through the enforcement of a protective distance or proper power and force limiting.
This is critical in \CAISs since human operators are expected to trust on ML-driven decisions even with limited or absent objective feedback on the ML performance in a specific production setting.
Indeed, there exist increasing concerns on how can people determine how much they could rely on ML models and their decision-making in safety critical domains~\cite{LuCHI2021}.

%
Risk management has been playing an important role in engineering processes tailored to robotic software and systems.
A common risk assessment approach for collaborative robotics applications is based on ISO 12100~\cite{iso201012100}.
Here, risk analysis is typically applied to manage requirements from elicitation to verification~\cite{cailliau2012probabilistic,felderer2014risk}.
However, the investigation of risk management approaches that take into account ML components embedded into robotic systems is still immature~\cite{ashmore2021ACMComputSurv}.
%
Over the last years, we have seen a proliferation of offline testing approaches for ML components as well as robustness assessment through adversarial perturbations~\cite{Zhang2020TSE}.
Recent empirical studies found offline approaches less effective in uncovering safety violations compared to online approaches~\cite{UlHaq2020Comparing}.
Furthermore, existing approaches used to generate adversarial perturbations mostly ignore the semantics and context of the overall system containing the ML component, whereas understanding the target world of interest plays a crucial role in this line of research~\cite{Gambi2019ISSTA}.
%
Considering ML in a closed-loop setting with the surroundings enables accurate and quantitative reasoning on the execution context of \CAISs as well as uncertainties and risks.
Nevertheless, ML components yield challenges not completely addressed by existing approaches as they are both hard to explain and sometimes have unexpected behavior due to their non-linear nature.

\subsection{Looking forward}

In the following, we list selected open issues. 
We believe these issues are urgent in light of the key characteristics of \CAISs. 

\emph{Risk assessment}. Existing standards, norms, and regulations of industrial and collaborative robots~\cite{iso51330,iso41571,iso62996} do not specifically describe how to deal with ML and related risks in \CAISs.
%
On the one hand, the robotics domain requires specific actions to justify the design and the implementation as an important part of system certification.
On the other hand, common techniques used in ML-enabled systems pose new challenges that require specific approaches in order to ensure adequate testing and validation as described by the latest standard documents on AI~\cite{isoTR24029,iso79799}.
The problem of integrating these latter techniques in risk assessment processes needed to certify robotic systems is still open.
{A}dequacy criteria for ML components is still an open issue. 
Furthermore, risk adequacy criteria for ML components is an open area.
Approaches based on neuron coverage have been proposed, recent findings indicate that this approach is not effective to spot real-world issues~\cite{Harel2020ESECFSE}.


\emph{Semantically meaningful assurance cases}. Researchers have shown that ML components can be easily fooled by generating unexpected inputs by using adversarial perturbation, neuron coverage driven, or mutation driven generation approaches.
However, deriving conclusions from the execution of such techniques is hard since inputs are often unrealistic/invalid since they do not take into consideration the world of interest and its semantics.
As highlighted by ISO/IEC DTS 4213 standard~\cite{iso79799} on assessment of ML classification performance, the problem of generating meaningful execution scenarios from suitable assurance cases that take an holistic perspective on both the system itself and the target world of interest is still open.

\emph{Acceptable behavior}. The definition of what is acceptable in terms of behavior of an ML component is not straightforward.
We still miss systematic approaches able to associate ML misbehavior to system-level faults.
Assuring the compliance of a human-robot collaboration requires us to specify the participation in the shared phenomena between the world and the machine. Proper techniques to model ML components along with their context to verify semantically meaningful properties are highly demanded.
It is worth noting that, as stated by the ISO/IEC TR 24029-1~\cite{isoTR24029}, robustness or more in general assessment of quality properties of ML components is an open area of research, and there are limitations to both testing and validation approaches.


\emph{Human aspects}. Human agents are a key part of \CAISs.
Even for simple applications, there is a substantial lack of information, especially before the production phase.
Complex, discontinuous and unforeseen collaboration scenarios may arise (e.g., different human operators, each one with different attitudes). 
This might affect the relevance of some of the features, composing the input space of the ML components, that can even change over time depending on the human reliance on ML-driven decisions.
Some modeling techniques tailored to human-in-the-loop systems have been recently introduced~\cite{Huck2020SPCE}. However, such approaches do not fully address variability and uncertainty in human behavior.
%

\emph{Understanding hazards}. Hazardous circumstances, such as unsafe situations for human operators induced by an ML component, shall be fully understood. 
However, there exists a striking lack of systematic engineering methods able to extract semantically meaningful information from ML misbehavior.
Explainable AI (XAI)~\cite{BARREDOARRIETA202082,HolzingerMLKE2021} is an open research area that can play a big role here. 
Indeed, it promises to guide developers in maintenance activities, such as retraining or feature re-engineering.
It also helps in increasing human reliance by explaining failures during validation and verification and thus mitigating compliance risks in production.

\section{Risk-driven Assurance of CAISs}

Our ongoing research~\cite{CamilliREFSQ2021,Camilli2021WAIN} aims at dealing with some of the major open issues reported above.
%
We believe that following a risk management perspective represents the key to monitor and control the likelihood and impact of undesired events, thus favoring the realization of opportunities promised by the usage of ML components in \CAISs.

\subsection{Approach Overview}

\begin{figure}[tb]
    \centering
    \includegraphics[width=0.5\textwidth]{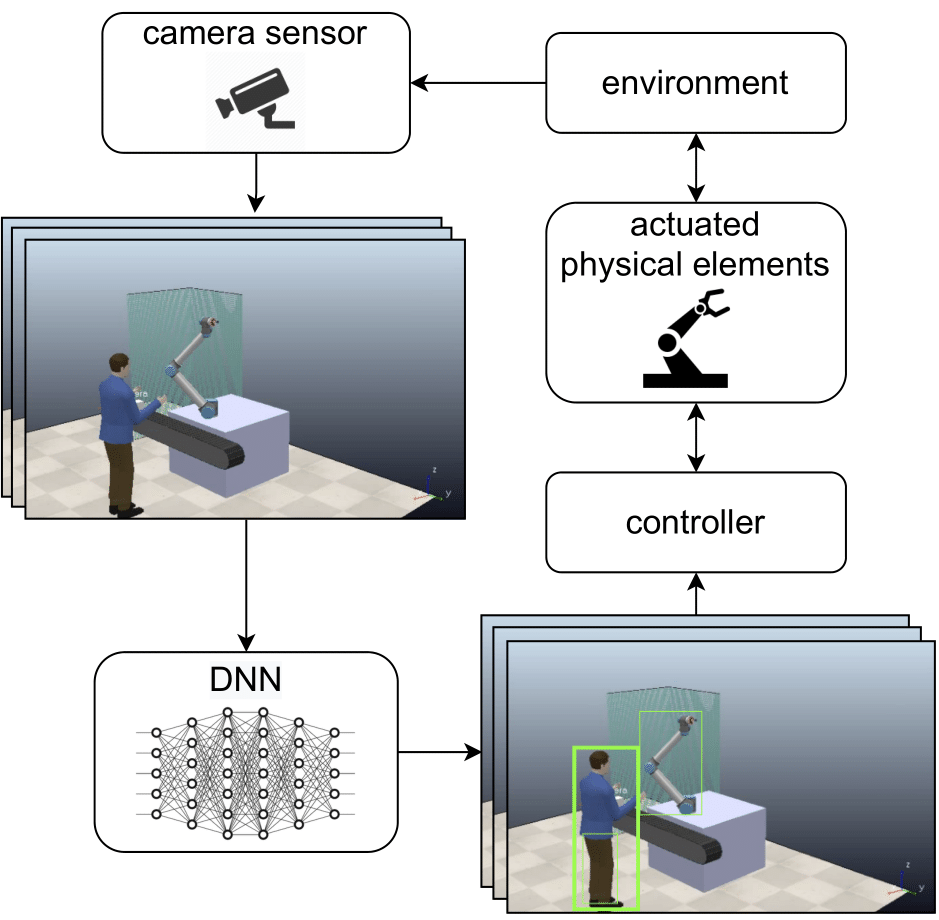}
    \caption
    {DNN visual perception component used to feed the controller of a robotic arm collaborating with a human operator.}
    \label{fig:cais_schema}
\end{figure}

Product-level standards for the safety of industrial robots are ISO 10218-1~\cite{iso51330} and ISO 10218-2~\cite{iso41571}.
These safety standards formally introduced a number of types of collaborative operation, such as \emph{safety-rated monitored stop}, \emph{speed  and  separation monitoring}, and \emph{power and force limiting}.
%
%
In all the three types of operation reported above, pre-collision control methods can exploit ML components able to perceive the surroundings of the robot. The ML component feeds a controller that in turn enacts the proper actions through the actuated physical elements.
For instance, let us consider an automatic emergency braking mechanism implemented by using a DNN visual perception component, as shown in Fig.~\ref{fig:cais_schema}.
In this case, the risk assessment of the collaborative operation aims at preventing any physical contact between the human operator and the moving robot.

\begin{figure*}[tb]
\centering
\begin{subfigure}{.55\textwidth}
  \centering
  \includegraphics[width=\linewidth]{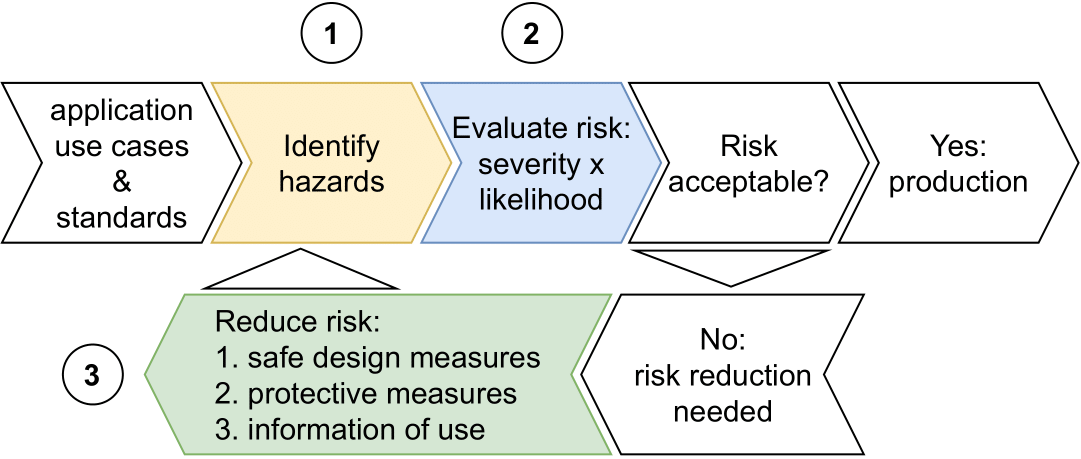}
  \caption{Risk assessment process ISO 12100.}
  \label{fig:iso_process}
\end{subfigure} \\
\begin{subfigure}{\textwidth}
  \centering
  \includegraphics[width=\linewidth]{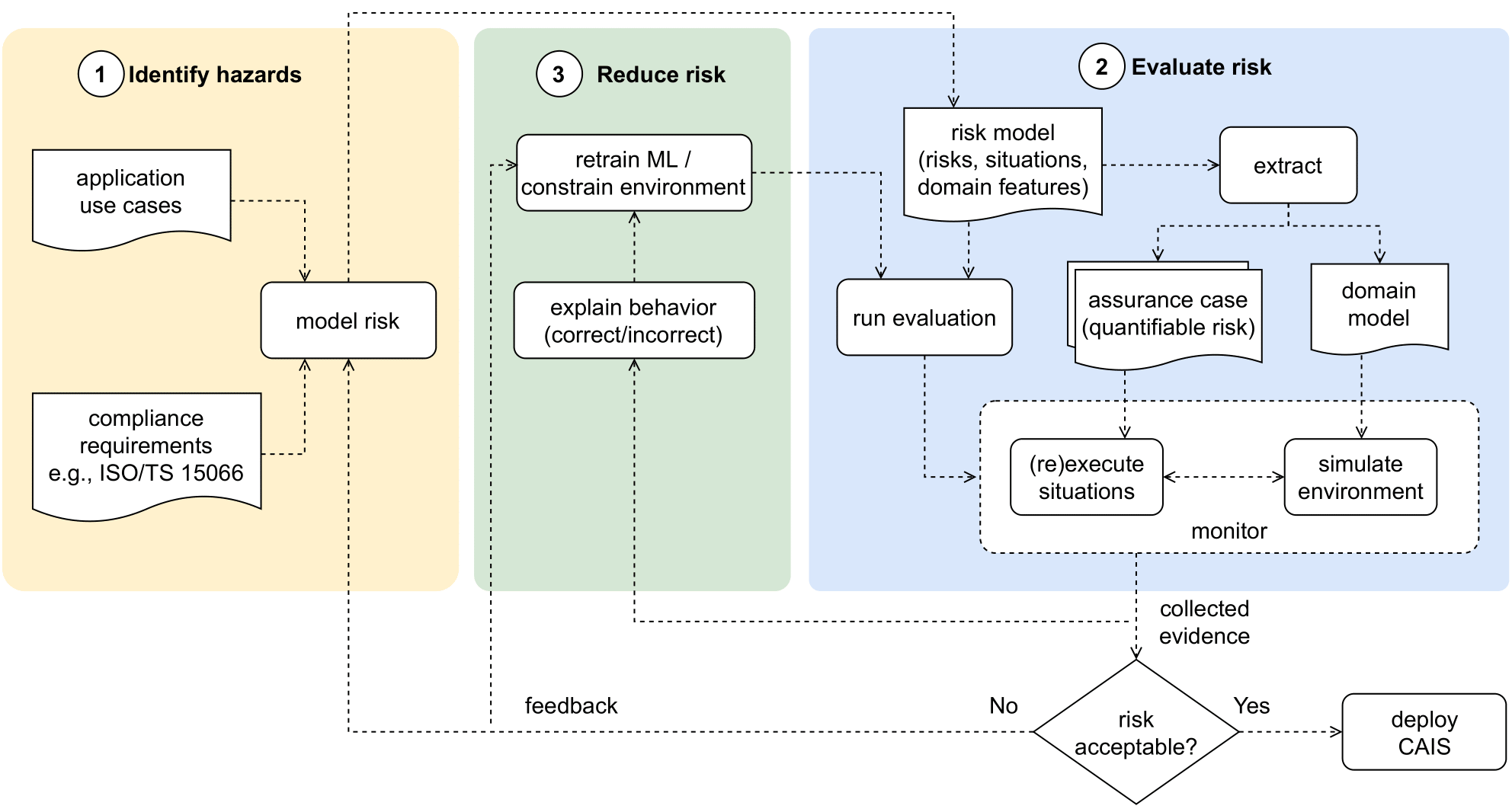}
  \caption{Mapping of the ISO 12100 to CAIS compliance assurance.}
  \label{fig:our_process}
\end{subfigure} \\
\begin{subfigure}{.5\textwidth}
\centering
    \includegraphics[width=\linewidth]{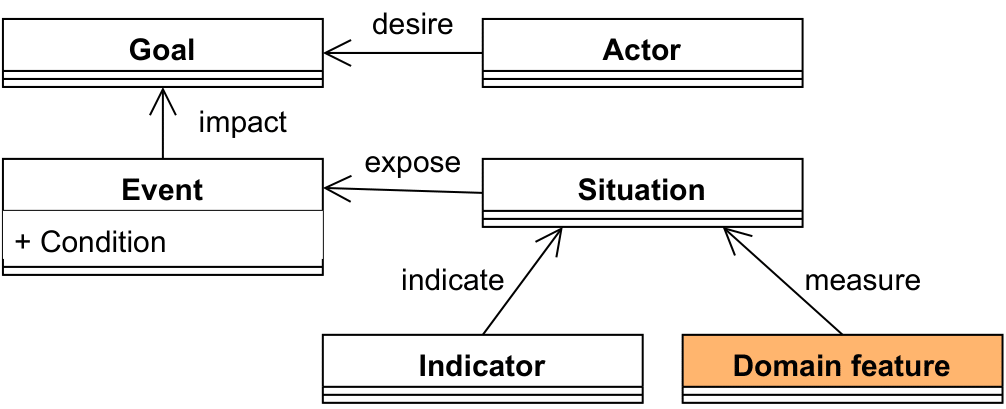}
    \caption{Excerpt of RiskML metamodel.}
    \label{fig:metamodel}
\end{subfigure}
\caption{
High level view on our risk-driven process and the extended RiskML metamodel.}
\label{fig:processes}
\end{figure*}

Figure~\ref{fig:iso_process} illustrates a typical risk assessment process extracted from the ISO 12100 standard~\cite{iso51528} regulating our domain of interest, i.e., collaborative applications.
The process aims at: (1) identifying the hazards; (2) evaluating; and then (3) reduce possible risks for each use case.
Figure~\ref{fig:our_process} zooms in the three aforementioned typical phases in the context of risk assessment and compliance assurance tailored to \CAISs.
In the step (1), all relevant application use cases involving potential hazards (e.g., physical contact between the operator and the robot) must be considered and specified. 
The specification shall consider the goals of the collaborative application, the relevant compliance requirements, as well as all possible \emph{situation}s that may break them.
In the step (2), all the situations and identified risks must be transferred to a number of assurance cases. 
Each assurance case shall provide structured argument supported by evidence to justify that the \CAIS, in a given situation, is assured with respect to a concern of interest (e.g., safety) in the intended operating environment.
In case empirical evidence is required, it may be collected by direct measurement of the phenomena of interest in production.
Nevertheless, given the safety critical nature of \CAISs, such measurements usually rely on simulators.
In case the collaborative application violates the compliance requirements, the collected evidence shall be used to trace violations back to key factors that affected negatively the behavior of the embedded ML component.
%
%
This information shall be used to apply risk reduction in the step (3).
For instance, misclassifications of the visual perception component can be explained in terms of physical characteristics of the environment that this component perceives.
Explanations can be used to retrain the perception module or further constrain the production environment.
%

\subsection{Identify hazards}

Our investigation aims at exploring suitable methods for risk representation and management tailored to \CAISs.
We recently introduced a risk representation method based on the RiskML language~\cite{DBLP:conf/er/SienaMS14} extended with the notion of \textit{domain feature} to model aspects of the environment that are perceived by an ML component and thus affect its behavior. 
Figure~\ref{fig:metamodel} shows an excerpt of the RiskML metamodel 
with the \emph{domain feature} extension (shown in orange). 


In the metamodel, the concept of \textit{situation} models the circumstances
under which a certain risk holds.
A situation could be, for example, a \emph{close collaboration} with the operator's hands in the shared physical space (e.g., upon the conveyor belt). A situation exposes to events. 
An \textit{event} represents a change in the circumstances affecting the system objectives and its occurrence is triggered by a \textit{condition}. 
Events have a certain positive (e.g., successful stop before collision) or negative (e.g., collision detected)
impact on the goals, and they occur with a certain likelihood. 
The notion of \textit{goal} models the desire of a stakeholder, modeled by the concept of \textit{actor}, to obtain or maintain some business value.
The concept of \textit{indicator} embodies a characteristic of a situation.
The concept of \textit{domain feature} enriches the RiskML meta-model by capturing relevant and quantifiable characteristics of the environment affecting the behavior of the ML component.
The domain feature space is a compact representation of all possible configurations under which the overall closed-loop \CAIS is expected to serve its purpose.
For instance, when analyzing the visual perception component of the robotic arm in Fig.~\ref{fig:cais_schema}, we may define a domain feature space consisting of the position and other physical characteristics of the human operator and the surrounding environment.
Even the position of the camera sensor can be a domain feature since it affects the field of view impacting the quality of the data sensed from the environment and therefore, the accuracy of the ML.
It is worth noting that the domain feature space has a different level of abstraction compared to the low-level feature space of pixels sensed by a camera and then analyzed by the ML, as shown in Fig.~\ref{fig:cais_schema}.
This makes the domain feature space better suited to model and then control simulated scenarios that can be easily rendered into meaningful inputs to the ML (i.e., images with specific pixel values) preserving the semantics of the world of interest.


%
%

\begin{example}[Domain feature]
\label{domain-feature-example}
Considering the CAIS in Fig.~\ref{fig:cais_schema}, a possible domain feature is the \emph{illuminance} of the environment.
Such a feature characterizes the production environment and can affect the ability to detect the presence of the human operator and then actuate the braking mechanism.
Its quantification 
typically depends on domain knowledge of the modeler. For instance, it can range over discrete categories 
or continuous values (e.g., between 100 and 10000 lux).
Other examples 
are the color of the background and color contrast between the background and the operator.
\end{example}

%

\subsection{Evaluate risk}

As shown in Fig.~\ref{fig:our_process}, the evaluation of risk starts by conveying 
elements from the risk model to compelling \emph{assurance case}s 
 relating together claims, evidence, and arguments.
Relevant assurance cases are mechanically derived from the RiskML model.
Specifically, the claim that a goal is achieved follows from the claim that the risk associated with the negative events is acceptable in a given situation.
The evidence supporting this latter claim shall be collected empirically as suggested by the standard ISO/IEC TR 24029-1~\cite{isoTR24029} by using a posteriori testing method that takes a holistic perspective to track the behavior of the ML component embedded into the whole system within a closed-loop simulated environment.
This allows the negative events to be detected by verifying the corresponding condition as defined in the RiskML model.
The condition formalizes a compliance requirement extracted from domain standards and it constrains measurable quantities characterizing the visible manifestation of the ML-driven behavior of the \CAIS.

\begin{example}[Event condition]
\label{event_example}
The negative event \emph{insufficient distance} defines a quantitative condition on the protective distance 
{(or safety zone)}
between the human and the robot as specified by the ISO/TS 15066 standard.
The misbehavior of the ML visual perception may affect the pre-collision controller and then propagate to measurable phenomena at system-level that yield hazardous circumstances. 
\end{example}

According to our process in Fig.~\ref{fig:our_process}, assurance cases extracted from the risk model are reproduced in a simulator of choice multiple times based on corresponding indicators and the domain feature space. 
%
%
Once the domain and range of the domain features are defined by the modeler, the next step is to find simulations in which negative events under consideration are likely to happen and therefore increase the risk.
Since exhaustive enumeration of all possible contexts is in general too expensive, our approach currently leverages meta-heuristic optimizing search techniques~\cite{10.5555/3001602} that use the feedback from the simulator to guide the sampling towards violations by either minimizing or maximizing the quantitative property expressed by the event condition (e.g., human-robot distance).
In the future we also plan to integrate and compare the cost-effectiveness of other testing methods, e.g., based on passive sampling including uniform random sampling or simulated annealing.
\begin{figure*}[tb]
\centering
\begin{subfigure}{0.7\columnwidth}
  \centering
  \includegraphics[width=\linewidth]{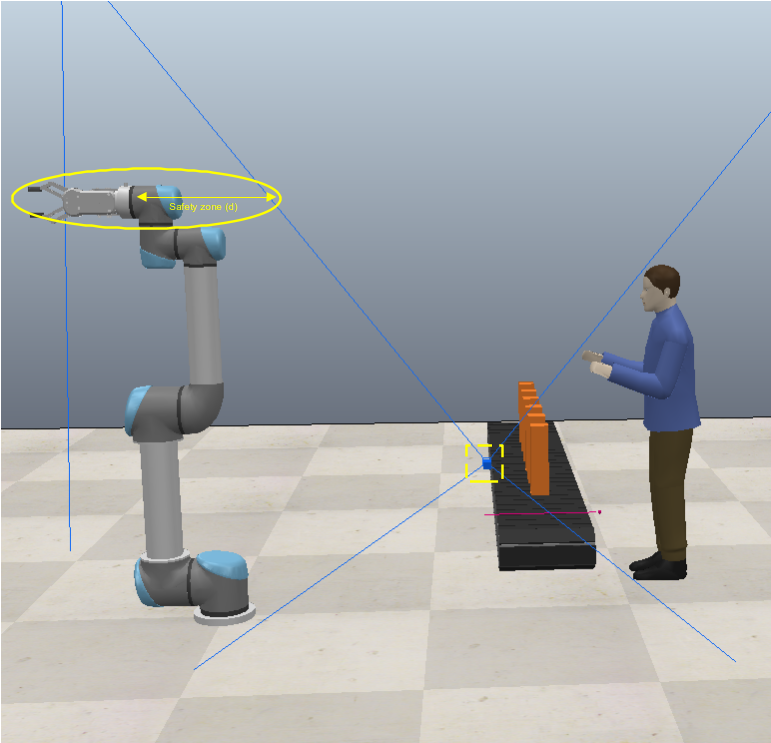}
  \caption{Safe collaboration example}
  \label{fig:sim1}
\end{subfigure}
\begin{subfigure}{0.7\columnwidth}
  \centering
  \includegraphics[width=\linewidth]{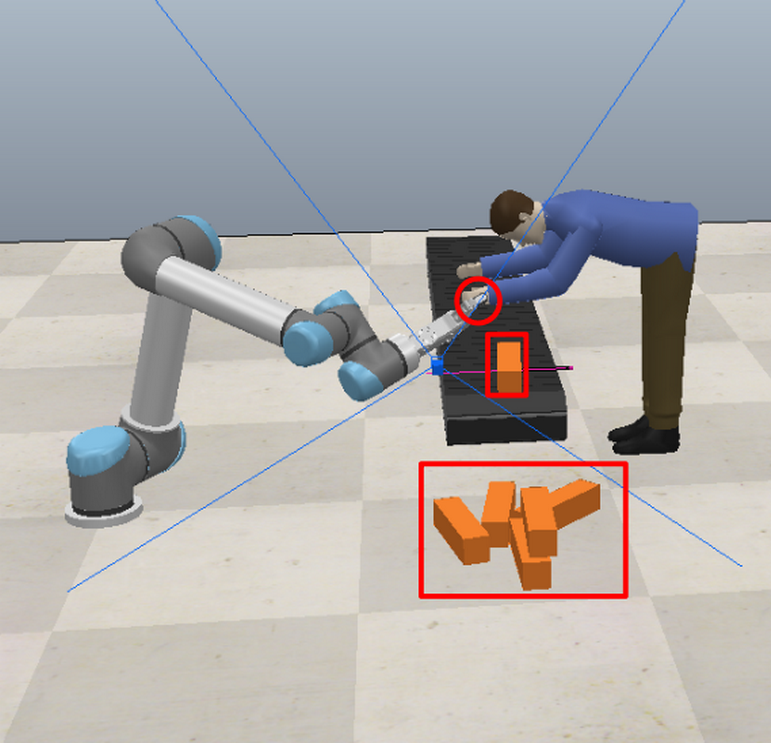}
  \caption{Unsafe collaboration example}
  \label{fig:sim2}
\end{subfigure}
\caption{Examples of situations executed by using our framework.}
\label{fig:simulation}
\end{figure*}



Our framework for prototyping and automated simulation-based testing is currently implemented by exploiting the robot simulator \textsc{CoppeliaSim}~\cite{coppeliaSim}.
Figure~\ref{fig:simulation} shows two images rendered from a simulated situation used to evaluate the risk of the event \emph{insufficient distance} (Example~\ref{event_example}).

The situation includes a robotic arm armed with a gripper, a conveyor belt, a set of pre-loaded objects on the conveyor belt and a human operator.
The blue wireframe shows the field of view of the vision sensor
(\textit{located within the dashed yellow frame})
that in turn feeds the DNN perception component trained to detect the moving objects and the hands of the operator.
Figure~\ref{fig:sim1} shows a safe collaboration according to the relative event condition, i.e., violation of the safety zone (
visualized as a yellow circle that represents a uniform distance around the robotic arm
).
Figure~\ref{fig:sim2} shows a safety violation that occurred while running the simulation multiple times driven by the testing process.
The optimizing search controls the execution context of the simulations by interacting with \textsc{CoppeliaSim} via a remote API.
The hazard detected in this example is caused by a change of the domain feature ``conveyor belt  speed'' from $0.1$ m/s to $0.5$ m/s.
This change causes the objects as well as the robotic arm to move faster, thus causing in turn an increase of the safety zone radius.
A timing mismatch between the objects and the actuated robotic arm causes the last object to fall off at the end of the conveyor belt.
Furthermore, as shown in the snapshot, even though the left hand of the operator is very close to the robotic arm, it is partially hidden from the field of view of the vision sensor.

\subsection{Risk Reduction}
\label{risk_reduction}
\begin{figure*}[tb]
    \centering
    \includegraphics[width=0.9\textwidth]{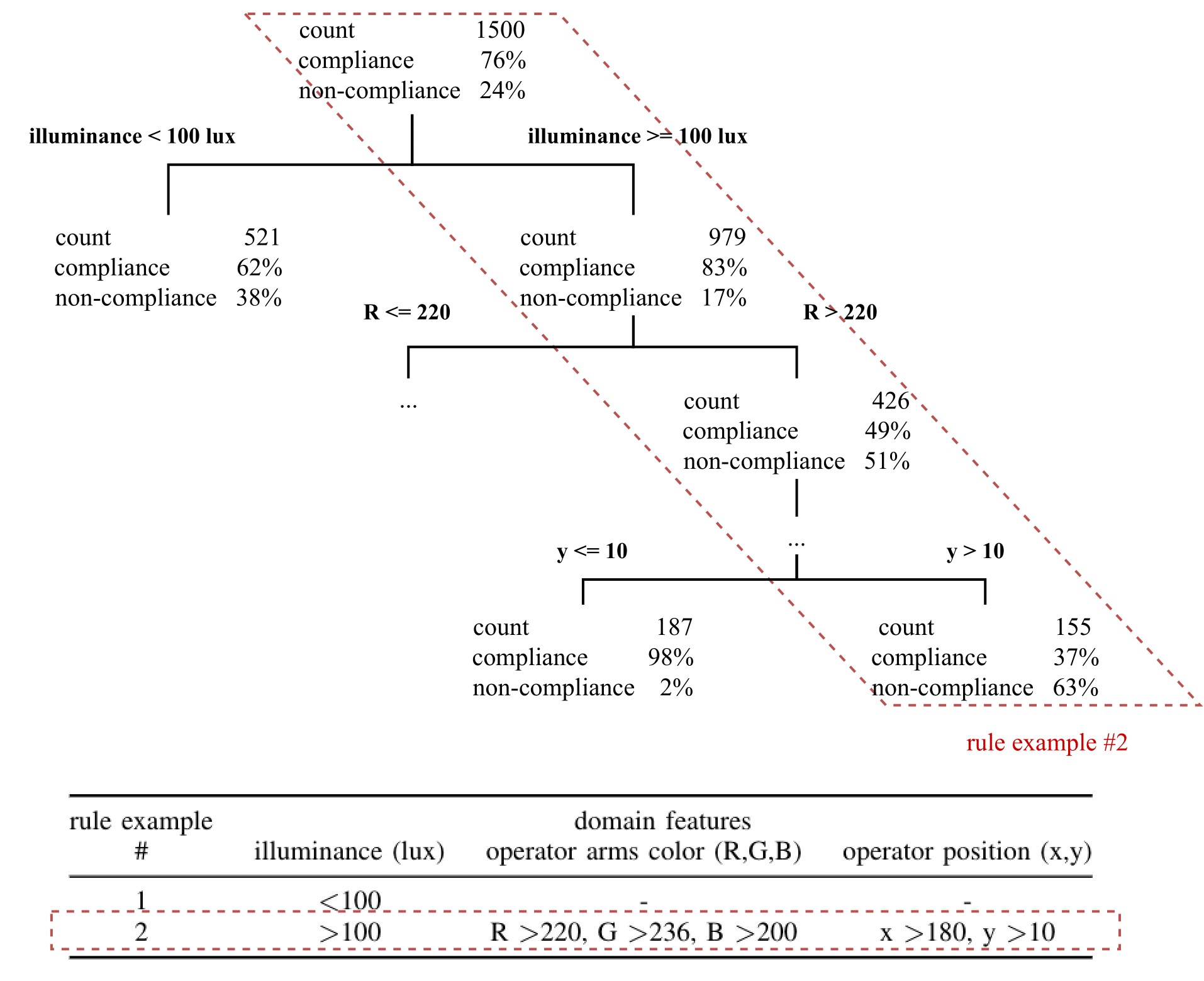}
    \caption{
    Example of decision tree and rule extraction.}
    \label{fig:tree}
\end{figure*}

According to the ISO/TS 15066 
{document}
on collaborative industrial collaborative robotics, the most preferred option is the elimination of the risk by applying a redesign or modification of (part of) the target system.
Here, we are particularly interested in the explainability of the ML components in terms of human-understandable characteristics of the situations that yield hazards.
Our approach aggregates the information extracted from multiple testing runs to automatically estimate the likelihood of a misbehavior and explain it in terms of human-understandable characteristics of the simulations.
%
%
Since the domain features are the relevant characteristics of a given simulation used to render meaningful inputs for the ML component, we use the actual value of these features to explain both acceptable and unacceptable behaviors observed during testing.
Each simulation can be labeled as either ``compliance'' or ``non-compliance'' depending on the occurrence of the negative events specified by the risk model.
Out of the testing process, we generate a labeled data set that maps valid assignments of domain features to labels.
Then, we use off-the-shelf methods to extract rules from this data set and explain the violations through a \emph{decision tree}.
By inspecting the tree we mechanically extract rules by following the paths to leaves having ``non-compliance'' associated with non-negligible likelihood, as shown by the example in Fig.~\ref{fig:tree}.

\begin{example}[Rule extraction]
\label{rule-extraction-example}
Figure~\ref{fig:tree} shows an example of decision tree and rule extraction from a dataset collected by testing the \emph{close collaboration} situation.
Illuminance values less than 100 (i.e., a very dark environment) lead to non-compliance with very high likelihood.
In this case, engineers can decide to reduce the risk by constraining the production environment and ensuring enough luminous flux incident on the shared working space.
With \emph{illuminance} greater than 100, there exist specific cases that increase the likelihood of misbehavior. Engineers may reduce the risk by retraining the DNN augmenting the training set with counterexamples built by following the rule \#2.
\end{example}

The automated computation of the decision tree closes the compliance assurance loop of the process in Fig.~\ref{fig:our_process} since it provides a feedback to the initial stage ``identify hazards'' by augmenting the risk model with a quantitative estimation of the likelihood associated with the negative events.
Furthermore it provides the third stage ``reduce risk'' with a counterexamples guided augmentation scheme, in which the set of rules, constraining the domain feature space, explain the misbehavior and thus can drive a retraining process of the ML component.




\section{Research Opportunities}

While the process of integrating ML components into critical applications like collaborative robotics is steadily growing, the scale impact of these systems is still little.
Assurance of these systems is an active research area that has produced a number of contributions especially focused on offline testing and adversarial robustness analysis of ML components.
Despite these contributions, much remains to be done to support the assurance process of \CAISs through rigorous, systematic, and disciplined approaches that take into account regulatory aspects. 
In our experience, significant work is required to improve the integration of verification techniques as defined by the standards documents on AI and the standards regulating the robotics domain.
To address the key challenges encountered or foreseen in our own work, we believe that software and systems engineers have to consider the quality of the ML components more holistically.
This means that the competing demands and attractions of ML itself and the world in which ML is employed must be appropriately balanced.
To achieve proper balance, we need sometimes to 
revert inclinations that reflect into current research trends.
Specifically, we have recently seen a proliferation of assurance methods that consider ML components offline. Thus, neglecting the dynamics of the world of interest, with which such components are deeply interleaved.
Emerging results demonstrate that offline approaches cannot be reasonably used to provide strong assurances for CAIS that shall instead be constructed around the goals of the collaborative applications, by respecting human needs and by assuring a safe collaboration online in a closed-loop setting with the surroundings.
%
Following this vision, we are currently exploring methods able to assure the integration of ML components by taking into consideration opportunities and risks emerging from the analysis of the world of interest of the collaborative application.
We are currently dealing with open challenges not completely addressed by existing approaches with the ultimate goal of increasing the scale impact of \CAIS in real-world settings.

We argue that the interaction between robotic agents equipped with ML components
and humans in critical collaboration scenarios is far from being considered safe, and in compliance with norms and regulations, with strong assurances.
The long term open question concerns how to engineer ``intelligent'' systems that collaborate with humans and enhance them by balancing the opportunity of compensating their weaknesses and the risk of compromising their well being by taking into account ethical, legal, and societal aspects~\cite{hagendorff2020ethics}.
To achieve advances, the research community shall collectively contribute towards a bidirectional understanding.
On the one hand, the behavior of human actors shall be understood by machines even in discontinuous and unforeseen collaboration scenarios, where tasks and preferences of humans as well as human reliance on ML-driven decisions can vary over time.
We need to address the limitations of current human modeling techniques and provide guarantees about control methods that take explicitly into account such models.
On the other hand, we need to build machines that are interpretable or explainable at the right level of abstraction to be understood by humans.
The explanation shall provide insights on the shared phenomena between the machine and the external world of interest in order to understand how to reduce the risks for human beings and ultimately build a high-trust workplace for them.

\section{Conclusion}
We reflected on the current state and our vision on compliance assurance of CAISs, where humans and AI systems share a physical space and work together to achieve a common goal even in safety critical settings. 
This motivated us to explore a disciplined risk-driven assessment process that takes into account regulatory aspects by integrating both domain standards and standards on AI.
Our current research activity focuses on three main directions; (1) compliance requirements extraction, representation, and traceability; (2) exploration of alternative hypotheses of risk representation and management for \CAISs; and (3) developing proper assurance methods which focuses on possible risks emerging from ML-driven decisions.
The ultimate goal of our multidisciplinary research team is to understand the right methods to reduce the risk of negative events affecting human beings and ultimately build a high-trust workplace for them.
Our research directions naturally emerged from the analysis of the current research landscape we conducted by looking backward at relevant topics that have been addressed and by looking forward at open issues and research opportunities that we expect to be addressed in future by the software and systems engineering research communities.
We believe that addressing these issues has broad implications on the scale impact of the usage of AI in critical domains.

\section{ACKNOWLEDGMENT}
This work was partially supported by the Austrian Science Fund (FWF): I 4701-N.

\bibliographystyle{IEEEtran}
\bibliography{references}

\begin{flushleft}

\begin{justify}
\textbf{Jubril Gbolahan Adigun} is a researcher and PhD candidate at the University of Innsbruck, Austria. Jubril holds a bachelor's degree in Electrical and Electronics Engineering from the University of Ilorin, Ilorin, Nigeria and a joint MSE degree in Software Engineering from the University of Tartu and Tallinn University of Technology, Estonia. His PhD focuses on the assurance of AI systems under uncertainty. Contact him at \href{mailto:jubril.adigun@uibk.ac.at}{jubril.adigun@uibk.ac.at}.
\end{justify}

\begin{justify}
\textbf{Matteo Camilli} received the PhD degree in Computer Science from the University of Milan in 2015. He is currently an assistant professor at the Free University of Bozen-Bolzano. His research activity focuses on formal methods and software engineering. He is especially interested in methods and tools to improve dependability of service-based, autonomous, cyber-physical, self-adaptive, and AI-based systems. He is part of the organizing committee of international conferences, such as IEEE ICSA (2022) and the track on software architecture of SIGAPP/ACM SAC (2020, 2021, 2022). Contact him at \href{mailto:matteo.camilli@unibz.it}{matteo.camilli@unibz.it}.
\end{justify}

\begin{justify}
\textbf{Michael Felderer} is a professor at the University of Innsbruck, Austria and a guest professor at the Blekinge Institute of Technology, Sweden. He holds a PhD and a habilitation degree in computer science. He has published more than 150 papers and received 12 best paper awards. His research interests include system quality and testing, AI and software engineering, as well as empirical software engineering. Further information can be found at \url{mfelderer.at}. Contact him at \href{mailto:michael.felderer@uibk.ac.at}{michael.felderer@uibk.ac.at}.
\end{justify}

\begin{justify}
\textbf{Barbara Russo} is full professor in Computer Science at the Free University of Bozen-Bolzano. She holds a PhD in pure mathematics from the university of Trento, Italy. She was visiting researcher at the Max Plan Institute for Mathematics, Bonn, Germany. She published more than 120 articles in pure mathematics and computer science. She is reviewer for the most relevant journals and conferences in software engineering and an associate editor of the Journal of Information and Software Technology. Her research interest is in software system engineering. Further information can be found at \url{www.inf.unibz.it/~russo}, Contact her at \href{mailto:barbara.russo@unibz.it}{barbara.russo@unibz.it}.
\end{justify}

\begin{justify}
\textbf{Andrea Giusti} is researcher and head of the Robotics and Intelligent Systems Engineering unit at Fraunhofer Italia Research, Italy. He received his Ph.D. degree in Robotics in 2018, from the Technical University of Munich (TUM), Germany. He received his Masters Degree in Mechatronic Engineering, summa cum laude, in 2013, and his Bachelors Degree in Telecommunications Engineering in 2010, both from University of Trento, Italy. His research interests include modelling and control of mechatronic systems, modular and reconfigurable robots, and human-robot collaboration. Contact him at \href{mailto:andrea.giusti@fraunhofer.it}{andrea.giusti@fraunhofer.it}.
\end{justify}

\begin{justify}
\textbf{Angelo Susi} is a research scientist and head of the Software Engineering unit at Fondazione Bruno Kessler in Trento, Italy. His research interests are in the areas of requirements engineering, software testing, and search-based software engineering. He published more than a hundred refereed papers in journals and international conferences and participated in the organization committee of conferences, such as SSBSE'12 (general chair), REFSQ (workshop and industry chair), RE (financial chair) and in program committees of international conferences and workshops. Contact him at \href{mailto:susi@fbk.eu}{susi@fbk.eu}.
\end{justify}

\begin{justify}
\textbf{Anna Perini}, FBK Distinguished Fellow,  former senior researcher at the Software Engineering research unit of Fondazione Bruno Kessler, Trento (Italy). Anna Perini teaches Requirements Engineering at the University of Trento, MSc degree in Computer Science. Her research interests include requirements engineering, agent-oriented software development methodologies, conceptual modelling, decision making in requirements engineering, and empirical studies. H-Index 43 (Google Scholar October 2021). Anna Perini served as program co-chair of the Int. Conference RCIS 2021, IEEE RE’19, REFSQ 2017, and of STAIRS 2006. She is member of the Steering Committee of the IEEE RE Int. conference and chair of the Steering Committee of the REFSQ conference. Moreover, she is serving as program committee member of several Int. Conferences, (e.g., ICSE, RE, CAiSE), and international workshops, and regularly reviews papers for top journals in the Software Engineering area. Contact her at \href{mailto:perini@fbk.eu}{perini@fbk.eu}.
\end{justify}

\begin{justify}
\textbf{Dominik T. Matt} holds the Chair for Production Systems and Technologies at the Faculty of Science and Technology at the Free University of Bozen-Bolzano where he also heads the research department ``Industrial Engineering \& Automation (IEA)''. He is the Director of the Fraunhofer Italia Research Center in Bolzano. 
He received his Ph.D. degree in Industrial Engineering from the Karlsruhe Institute of Technology (KIT). 
During his academic career, he has authored more than 250 scientific and technical papers in journals and conference proceedings; 
He is listed among the most cited professors in 2020 (top 100.000) as well as among the 2\% top scientists for long career-impact in their specific fields. He is a member of numerous national and international scientific organizations and committees. Since 2020 he is also a member of the renowned German National Academy of Science and Engineering ``acatech'' (\url{www.acatech.de}). His research focuses on the areas of Industry 4.0 and Smart Factory, the application of Artificial Intelligence in industrial production, Lean Production, as well as the planning and optimization of assembly systems and internal logistics. Contact him at \href{mailto:dominik.matt@unibz.it}{dominik.matt@unibz.it}. \end{justify}
\end{flushleft}
\end{document}